\documentclass[12pt]{article}

\usepackage{amsbsy,amsmath}
\usepackage{amsfonts}
\usepackage{amssymb}
\usepackage{amscd}
\usepackage{bbm}
\usepackage{fancybox}
\usepackage{cite}
\usepackage{amsmath,amsfonts,amsbsy}
\usepackage{pstricks,pst-node}
\usepackage[small,bf,hang]{caption2}
\usepackage{graphicx}
\usepackage{epsfig}
\usepackage{psfrag}
\usepackage{comment}

\numberwithin{equation}{section}
\begin{document}

\begin{titlepage}
\begin{center}

\vskip 1.5cm

{\Large \bf An interacting conformal chiral 2-form electrodynamics in six dimensions$^*$}

\vskip 1cm

{\bf Paul K.~Townsend} \\

\vskip 25pt

{\em $^2$ \hskip -.1truecm
\em  Department of Applied Mathematics and Theoretical Physics,\\ Centre for Mathematical Sciences, University of Cambridge,\\
Wilberforce Road, Cambridge, CB3 0WA, U.K.\vskip 5pt }

{email: {\tt P.K.Townsend@damtp.cam.ac.uk}} \\

\end{center}

\vskip 0.5cm
\begin{center} {\bf ABSTRACT}\\[3ex]
\end{center}

The strong-field limit for the 2-form potential on an M5-brane yields a conformal chiral 2-form electrodynamics in six dimensions, with gauge-invariant self-interactions but no adjustable coupling constant;  the stress tensor is that of a null fluid. Lorentz invariance  can be made manifest via an interpretation as a tensionless `space-filling M5-brane', or as a truncation of the 
infra-red dynamics of an M5-brane in AdS$_7\times S^4$.

\vfill

$*\ $ Contribution to a volume in celebration of Michael Duff's 70th year. 

\end{titlepage}
\tableofcontents

\section{Introduction}

In 1983, on the occasion of the 60th birthday of Jan \L{}opusza\'nski, Iwo Bialynicki-Birula wrote an article
about Born-Infeld (BI) electrodynamics in which he showed, {\it inter alia}, that there is a strong-field limit
that yields a conformal-invariant {\sl interacting} electrodynamics \cite{BialynickiBirula:1984tx}.  It was shown recently by Luca Mezincescu and the author \cite{Mezincescu:2019vxk} that many of the properties of this Bialynicki-Birula electrodynamics (BBE) follow from an interpretation of it as a truncation of the low-energy dynamics of a tensionless D3-brane. As we also showed, it has an alternative interpretation as a truncation of the infra-red (IR) dynamics of a D3-brane in the AdS$_5\times S^5$ vacuum of IIB supergravity. 

On the occasion of Michael Duff's 70th year (70 is the new 60) it is my pleasure to offer an extension of the ideas
in \cite{BialynickiBirula:1984tx,Mezincescu:2019vxk} that makes contact with two principal themes of Michael's research
over the last 40 years. One is Eleven Dimensions and the other is Branes;  Michael's recent review \cite{Duff:2018goj} 
reminds us that the two topics merged about 30 years ago, only to meet with ``fierce resistance''. 

The starting point here will be the M5-brane of M-theory, and more specifically the dynamics of its worldvolume 2-form potential with self-dual 3-form field-strength. This 2-form potential is part of a (2,0) supermultiplet in six dimensions (6D) that couples to the boundary of a membrane \cite{Townsend:1995af}. Membranes stretched between parallel M5-branes become tensionless strings in the coincidence limit
\cite{Strominger:1995ac} and the IR limit of the resulting dynamics is some interacting (2,0)-supersymmetric 6D conformal field theory (CFT) \cite{Witten:1995zh}. This (2,0) theory remains ``mysterious'', in part because it has no coupling constant and hence cannot be approached perturbatively from some free-field limit;  it has been argued that, away from the conformal limit,  it becomes  a free-field theory of a (2,0) supermultiplet coupled to tensile self-dual 6D strings \cite{Arvidsson:2004xa}.

The main result here is the construction of an interacting conformal chiral 2-form electrodynamics in six dimensions; this too has no coupling constant and (for reasons given at the end of this paper) is likely to have a (2,0)-supersymmetric extension. It is also 
related to M5-brane  dynamics, in the same way that BBE is related to D3-brane dynamics. It is unlikely (in the author's estimation) to be the (2,0) CFT  of multiple M5-brane dynamics but it may provide some useful insights into that problem.  Because of the close analogy to BBE it will  be useful to first consider the basic structure of BBE; its method of construction, some of its properties, and its relation to the 
D3-brane will be reviewed later. 

A feature of BBE emphasised in \cite{BialynickiBirula:1984tx} is that it has no configuration-space action. However, there {\sl is} a phase-space action; for standard Minkowski spacetime coordinates $\{t,\boldsymbol{\sigma}\}$ it is
\begin{equation}\label{BBact1}
S_{BBE} = \int dt\! \int\! d^3\sigma \left\{ {\bf E}\cdot {\bf D} - |{\bf D}\times {\bf B}|\right\} \, . 
\end{equation}
The (3-covector) electric field ${\bf E}$ and the (3-pseudovector density) magnetic-induction field ${\bf B}$  are defined in terms of the components $(A_0, {\bf A})$ of a potential 1-form in the usual way:
\begin{equation}\label{EandB}
{\bf E} = \boldsymbol{\nabla} A_0- \dot{\bf A} \, , \qquad {\bf B}= \boldsymbol{\nabla} \times {\bf A} \, . 
\end{equation}
Notice that 
\begin{equation}
{\bf E}\cdot{\bf D} = -\dot{\bf A} \cdot {\bf D} - A_0 \boldsymbol{\nabla}\cdot {\bf D} 
+ {\rm total\ derivative}\, , 
\end{equation}
which shows that $A_0$ is a Lagrange multiplier for the Gauss-law constraint on the electric-displacement $3$-vector density ${\bf D}$, which is canonically conjugate to $-{\bf A}$.  The BBE action is not obviously conformal invariant, and even its Lorentz invariance is not obvious, but these properties were established in \cite{BialynickiBirula:1984tx} and can be simply explained in terms of its D3-brane origin \cite{Mezincescu:2019vxk}. 

A comment on terminology may be useful here. Given a phase-space action of the form 
\begin{equation}\label{NLE}
    S= \int\! dt\! \int d^3\sigma \left\{ {\bf E}\cdot {\bf D} - {\cal H}(D,B)\right\} \, , 
\end{equation}
the magnetic field is defined as ${\bf H} = \partial {\cal H}/\partial {\bf B}$; it is a (pseudo) $3$-covector whereas the magnetic-induction field ${\bf B}$ is a (pseudo) $3$-vector density.  The Hamiltonian of Maxwell electrodynamics in a vacuum is such that ${\bf H}={\bf B}$ in cartesian coordinates (for which the distinction between vectors, covectors and vector densities is lost) and there is then no need to distinguish between ${\bf H}$ and ${\bf B}$. However,  the distinction is essential to the non-linear BI theory, and to BBE.  Similarly, the electric $3$-covector field ${\bf E}$ may be defined as ${\bf E} = \partial {\cal H}/\partial {\bf D}$, but this is just the field equation derivable from the phase-space action by variation with respect to ${\bf D}$. For the Maxwell case, and in cartesian coordinates,  this equation is ${\bf E}={\bf D}$, but equality no longer holds in the non-linear BI theory, or in BBE.  

The 2-form analog of BBE in six dimensions also has no configuration-space action but it too has a phase-space action. To present it we must first introduce analogs of the electric field and 
magnetic-induction field in terms of the components of the 3-form field-strength $F=dA$ for a 2-form potential $A$: 
\begin{eqnarray}
E_{ij} &=& F_{ij0} \equiv \dot A_{ij} - 2 \partial_{[i} A_{j]0}\, , \nonumber \\
B^{ij} &=& \frac16 \varepsilon^{ijklm}F_{klm} \equiv \frac12 \varepsilon^{ijklm}\partial_k A_{mn}\, . 
\end{eqnarray}
The phase space action involves an analog of the electric-displacement field (now an antisymmetric tensor density) with components
$D^{ij}$ that are canonically conjugate to $A_{ij}$ but there is also a chirality constraint that allows it to be eliminated; the
resulting action is a functional only of $A_{ij}$ (i.e. the space components of the 2-form potential $A$). This action is  
\begin{equation}\label{strongfield}
S[A]=  \int dt\! \int\! d^5\sigma \left\{ - \frac12\dot A_{ij} B^{ij} - |B\wedge B|\right\}\, ,
\end{equation}
where  $B\wedge B$ is a 5-vector density with components
\begin{equation}
(B\wedge B)_i = \frac14 \varepsilon_{ijklm}B^{jk}B^{lm}\, . 
\end{equation}

The main aim of the remainder of this paper will be to explain why this action is Lorentz invariant, and not only that but also conformal invariant. This will be achieved in two ways, both related to limits and truncations of the dynamics of an M5-brane, for which the manifestly Lorentz invariant, and reparametrization invariant, action was found in \cite{Bandos:1997ui}, although our starting point will be the phase-space action constructed from it in \cite{Bergshoeff:1998vx}. The simplest way is via a tensionless limit of the M5-brane in the 11D Minkowski vacuum, and this will be discussed first. Then we shall see how the same theory arises from an IR limit of an 
M5-brane in the AdS$_7\times S_4$ vacuum of 11D supergravity. Throughout the paper, except for some final comments and speculations, all fermions are set to zero.

As already mentioned, most of these M5-brane-derived results are generalizations of the D3-brane-derived results of \cite{Mezincescu:2019vxk}, and of the idea of a strong-field limit first explored in \cite{BialynickiBirula:1984tx}, so it will be very useful to cover some of that ground first.  It was brought to the author's attention after submission to the arXiv of the first version of this paper that 
some of this material is also covered in a paper of Gibbons and West \cite{Gibbons:2000ck}, who further consider a ``strong-coupling limit'' of the M5-brane with results that overlap with those 
found here.

\section{Born-Infeld preliminaries}

The Born-Infeld (BI) theory of non-linear electrodymamics has an action of the form
\begin{equation}\label{BIact}
S_{BI}[A]= -T \int\! d^4x \, \left\{ \sqrt{-\det(\eta + T^{-\frac12} F)} -1\right\} \, , 
\end{equation}
where $T$ is a constant with dimensions of energy density (in units for which $\hbar=1$), and $F=dA$ is the Faraday 2-form for the electromagnetic  1-form potential $A$. We shall need the corresponding phase-space action, which
takes the form (\ref{NLE}) with Hamiltonian density 
\begin{equation}\label{MongeBI}
{\cal H}_{BI} = T \left\{ \sqrt{1 + T^{-1} \left(|{\bf D}|^2 + |{\bf B}|^2\right) 
    + T^{-2}|{\bf D}\times {\bf B}|^2} -1\right\}\, . 
\end{equation}
Notice that $H_{BI}$ has an $SO(2)$ invariance under rotations of $({\bf D},{\bf B})$.  

In the weak-field limit we have 
\begin{equation}
{\cal H}_{BI} \to {\cal H}_{Maxwell} = \frac12\left( |{\bf D}|^2 + |{\bf B}|^2\right)\, .  
\end{equation}
Notice that the constant $T$ has dropped out; this is related, of course, to the conformal invariance
of Maxwell's equations. 

To investigate the strong-field limit it is convenient to rewrite the Hamiltonian density as
\begin{equation}\label{BIham}
{\cal H}_{BI} = \sqrt{|{\bf D}\times {\bf B}|^2 + T  \left(|{\bf D}|^2 + |{\bf B}|^2\right) +T^2} -T\, , 
\end{equation}
and then use the fact that the strong-field limit for fixed $T$ is equivalent to the $T\to0$ limit at fixed field strengths. This is precisely how Bialynicki-Birula found the Hamiltonian density 
\begin{equation}
{\cal H}_{BB}= |{\bf D}\times {\bf B}|\, . 
\end{equation}
The corresponding BBE field equations are 
\begin{equation}\label{BBeofm}
\dot {\bf D}  =   \boldsymbol{\nabla} \times ({\bf n}\times {\bf D} )\, , \qquad \dot {\bf B}  =  \boldsymbol{\nabla} \times ({\bf n}\times {\bf B} )\, ,  
\end{equation}
where ${\bf n}$ is the unit-vector field
\begin{equation}
{\bf n} = {\bf D}\times {\bf B}/ |{\bf D}\times {\bf B}| \, . 
\end{equation}
These equations should be taken together with the constraint $\boldsymbol{\nabla} \cdot {\bf D}=0$ and the identity 
$\boldsymbol{\nabla} \cdot {\bf B}=0$. As observed in \cite{BialynickiBirula:1984tx}, the equations are non-linear because only solutions with the same ${\bf n}$ can be superposed. 
Notice that the BBE Hamiltonian is independent of the dimensionful constant $T$, which suggests that BBE might be conformal invariant. In fact, it {\sl is}  conformal invariant, and it is also invariant under an enlarged $Sl(2;\mathbb{R})$  electromagnetic duality group \cite{BialynickiBirula:1984tx}. 

\subsection{Poincar\'e and conformal invariance}

The BI configuration-space action (\ref{BIact}) is manifestly Poincar\'e invariant. This symmetry is not obvious from the phase-space action with BI Hamiltonian density (\ref{BIham}) but the equivalence of this action to (\ref{BIact}) can be established by a straightforward elimination of the momentum variable ${\bf D}$. However, this step cannot be performed once the strong-coupling limit has been taken as the BBE phase-space action of (\ref{BBact1}) is {\it linear} in ${\bf D}$, and this puts into doubt the Poincar\'e invariance of BBE.  The issue was addressed and resolved in \cite{BialynickiBirula:1984tx} but there is another way to exhibit the spacetime  symmetries \cite{Mezincescu:2019vxk}; we review the idea here but taking the BI action as our starting point, rather than its Bialynicki-Birula limit.   

In writing down the BI action (\ref{BIact}) we implicitly made the standard choice of Minkowski coordinates for the Minkowski spacetime. Let us now rename these Minkowski coordinates as $\{X^\mu; \mu=0,1,2,3\}$; then, in any other coordinate system $\{\xi^\mu; \mu=0,1,2,3\}$, specified by the functions $X^\mu(\xi)$, the Minkowski spacetime metric is 
\begin{equation}
g_{\mu\nu}(\xi) = \partial_\mu X^\rho\partial_\nu X^\sigma \eta_{\rho\sigma}\, ,  
\end{equation}
where $\partial_\mu$ is a partial derivative with respect to $\xi^\mu$. The BI action is now
\begin{equation}\label{BIreparam}
S_{BI}[A,X] = -T\int \! d^4\xi \,\left\{ \sqrt{-\det(g+ T^{-\frac12}F)} - \sqrt{-\det g} \right\}\, . 
\end{equation}
This action is reparametrisation invariant; it depends on the four functions $X^\mu$ in addition to $A$, and this dependence must be taken into account when passing to the phase-space form of the action. After writing $\xi^0 =t$ and $\xi^i = \sigma^i$ ($i=1,2,3)$, this new phase-space action takes the form  
\begin{equation}\label{BIphase}
    S[{\bf A},X; {\bf D}, P; A_0, u]= \int dt \int\! d^3 \sigma\, \left\{ \dot X^\mu P_\mu  +  
    {\bf E}\cdot {\bf D} - u^\mu{\cal H}_{\mu}\right\}\, , 
\end{equation}
where $P_\mu$ is the momentum density canonically conjugate to $X^\mu$. The new Lagrange multipliers $u^\mu$ impose
an additional four constraints (i.e. in addition to the Gauss-law constraint); the constraint functions are
\begin{eqnarray}\label{BIcons}
{\cal H}_0 &=& \frac12\left\{ (P-TC)^2  + T \left( D^iD^j + B^iB^j\right)h_{ij} + T^2\det h \right\}\, ,  \nonumber \\
{\cal H}_i &=& \partial_i X^\mu P_\mu - \varepsilon_{ijk}D^i B^k\, , 
\end{eqnarray}
where $h$ is the space metric\footnote{The change of letter avoids a potential confusion of its inverse with the inverse of $g$.}; i.e. $h_{ij} = g_{ij}$, and the 4-vector-density $C$ has components 
\begin{equation}
C_\mu = \frac16 \varepsilon_{\mu\nu\rho\sigma} \, \varepsilon^{ijk} \partial_i X^\nu\partial_j X^\rho\partial_k X^\sigma\, .
\end{equation}
Useful identities are
\begin{equation}
C^2 \equiv -\det h\, , \qquad \partial_i X^\mu C_\mu \equiv 0 \, . 
\end{equation}

Before proceeding, we pause to explain how the BI Hamiltonian density of (\ref{MongeBI}) is recovered from the new, reparametrisation invariant, action. The key point is that the new constraints are first-class; more precisely, the constraint functions ${\cal H}_\mu$ 
form a first class set, which requires the Poisson-bracket of each element of the set with any other constraint function to be zero
on the surface in phase-space defined by the set of all constraints. Here it should be recalled that in addition to the new constraints 
imposed by the new Lagrange multipliers $u^\mu$ there is also the Gauss-law constraint imposed by $A_0$; this is first-class too
because its PBs with ${\cal H}_\mu$ are zero, as expected because it generates the gauge transformation of ${\bf A}$ and the functions ${\cal H}_\mu$ are gauge invariant. It follows that the new constraint functions ${\cal H}_\mu$ will form a first-class set if 
the matrix of their PBs is zero when all ${\cal H}_\mu$ are zero. This first-class property, which was verified in detail in \cite{Mezincescu:2019vxk}, implies that the new constraints also generate gauge invariances. These new gauge invariances are (on-shell) equivalent to diffeomorphisms of the 4D Minkowski spacetime, and this allows us to impose the Monge gauge: 
\begin{equation}
X^\mu(\xi) = \xi^\mu \quad (\mu=0,1,2,3)\, \quad \Rightarrow\quad h_{\mu\nu} =\eta_{\mu\nu}\, . 
\end{equation}
In this gauge $h_{ij}=\delta_{ij}$ ($\Rightarrow\ \det h=1$) and $P^\mu C_\mu =P^0$, so that 
\begin{equation}
    2{\cal H}_0 = -(P^0+T)^2 + |{\bf P}|^2 + T (|{\bf D}|^2 + |{\bf B}|^2) +T^2\, . 
\end{equation}
The solution of the constraints for $P_\mu$ is then 
\begin{eqnarray}\label{Pmu}
P^0 &=& \pm \sqrt{|{\bf D}\times {\bf B}|^2 + (|{\bf D}|^2 + |{\bf B}|^2) +T^2} - T\, ,\nonumber\\
{\bf P} &=&  {\bf D}\times {\bf B} \, . 
\end{eqnarray}
The Monge gauge Hamiltonian is $P^0$, which is ${\cal H}_{BI}$ if we assume $P^0>0$.  

Having now confirmed its equivalence to the BI action, we return to the reparametrisation-invariant action (\ref{BIphase}). 
Its advantage is that the Poincar\'e  group now acts {\sl linearly}, and {\sl only} on $(X,P)$.  The Noether charges corresponding to spacetime translations and Lorentz transformation are simply
\begin{equation}
{\cal P}_\mu = \int\! d^3\sigma\, P_\mu\, , \qquad 
{\cal J}^{\mu\nu} = 2\int\! d^5\sigma\, X^{[\mu} P^{\nu]}  \, . 
\end{equation}
It is straighforward to verify, using the $(X,P)$ equations of motion and the constraints,  that these charges are time-independent for appropriate boundary conditions, and that they span the algebra of the Poincar\'e  group with respect to the Poisson bracket relations derived from the action (\ref{BIphase}).  As Noether charges are gauge invariant, they are unchanged by the imposition  of the Monge gauge, except that the previously independent momentum variables $P_\mu$ must be replaced by their Monge gauge expressions (\ref{MongeBI}), and all PB relations must now be computed using the canonical PB relations of the Monge-gauge action. The Noether charges are still time-independent (now as a consequence of the Monge gauge field equations) and their PB algebra is also unchanged. What does change is that the Noether charges now generate transformations of the phase-space variables $({\bf A}, {\bf D})$, which were initially inert! 

All of this discussion continues to apply in the $T\to 0$ limit. In this limit the constraint functions (\ref{BIcons}) 
simplify to 
\begin{equation}
{\cal H}_0 = \frac12 P^2\, , \qquad {\cal H}_i = \partial_i X^\mu P_\mu - \varepsilon_{ijk}D^i B^k\, . 
\end{equation}
These are still Poincar\'e invariant but the action (\ref{BIphase}) is now invariant under the larger group of conformal isometries
of the 4D Minkowski metric.  By imposing the Monge gauge and solving the constraints for $P_\mu$ one recovers the 
BBE action (\ref{BBact1}), which is therefore also Poincar\'e, and conformal,  invariant. The Monge-gauge
expressions for the associated Noether charges can be used to find the symmetry transformations; this step was 
carried out for Lorentz transformations in \cite{Mezincescu:2019vxk}, and the result was used to perform a direct
verification of the Lorentz invariance of the BBE action of (\ref{BBact1}).

\subsection{Relation to the D3-brane}

We have now arrived at a point from which it is easy to see the relation of BBE to the D3-brane.
The action  (\ref{BIphase}) is a truncation of the phase-space action for the D3-brane of tension $T$ in the Minkowski vacuum of 10-dimensional IIB supergravity \cite{Bergshoeff:1998ha}: the transverse fluctuations of a static planar D3-brane 
are ignored (as are all anticommuting spinor variables) thereby reducing the 10D Minkowski spacetime to a 4D Minkowski spacetime that can be identified with the D3-brane worldvolume. As the brane now fills the available space this truncation was referred to as  ``space-filling''. The only remaining dynamics is that of the 1-form potential on the 4D Minkowski worldvolume; in general this is 
equivalent to the BI theory of non-linear electrodynamics,  with $T$ being its dimensionful constant. The tensionless limit 
is thus equivalent to the strong-field limit of \cite{BialynickiBirula:1984tx}. 

As already emphasised, the advantage of the D3-brane interpretation of BBE is that Lorentz invariance is linearly realized, and hence manifest.  Conformal invariance is still non-linearly realised but it is now related to the (easily established) conformal invariance of null brane dynamics. A further advantage of the D3-brane perspective is that it relates the $Sl(2;\mathbb{R})$ electromagnetic duality invariance group of BBE found in  \cite{BialynickiBirula:1984tx} to the $Sl(2;\mathbb{R})$ duality group of IIB supergravity \cite{Mezincescu:2019vxk}. As this latter $Sl(2;\mathbb{R})$ is broken to $Sl(2;\mathbb{Z})$ in IIB superstring theory, we may anticipate that quantum effects will break the $Sl(2;\mathbb{R})$ invariance of BBE to $Sl(2;\mathbb{Z})$. 

Finally, by considering a D3-brane in the AdS$_5\times S^5$ vacuum  of IIB supergravity one can interpret BBE as a truncation of the IR limit of the dynamics of a planar D3-brane coincident with the AdS$_5$ Killing horizon \cite{Mezincescu:2019vxk}. This suggests that one might find a chiral 2-form electrodynamics in six dimensions by taking the IR limit of a planar M5-brane in the AdS$_7\times S^4$ vacuum of 11-dimensional supergravity. As we shall see later, this is indeed one way of arriving at the chiral 2-form electrodynamics advertised in the Introduction. However, in complete analogy with the D3-brane case, it can also be found more simply as a truncation of the action for a tensionless M5-brane in the 11-dimensional Minkowski vacuum, and this will be our starting point for what follows.

\section{The M5-brane} 

A reparametrisation-invariant phase-space action for the M5-brane, in a general bosonic background, was obtained in \cite{Bergshoeff:1998vx} from the configuration-space M5-brane action of \cite{Bandos:1997ui}. Here we take the background to be the 11-dimensional Minkowski vacuum, 
with Minkowski coordinates $\{X^m; m=0,1, \dots,10\}$,  and we set to zero all fermions. The worldvolume coordinates $\{\xi^\mu; \mu=0,1, \dots,5\}$  are split into $\xi^0 =t$ and $\xi^i=\sigma^i$ ($i=1,\dots,5$). The action then takes the form 
\begin{equation}\label{M5}
S_{M5} = \int\! dt \int\! d^5\sigma \left\{ \dot X^m P_m + \frac12 \dot A_{ij} D^{ij} - u^\mu {\cal H}_\mu - \frac12\sigma_{ij} \chi^{ij} \right\}\, , 
\end{equation}
where the variables $u^\mu$ and $\sigma_{ij}$ ($= -\sigma_{ji}$) are Lagrange multipliers for constraints. The constraint functions ${\cal H}_\mu$ and $\chi^{ij}$ are functions of canonical variables that will be given below, we may ignore the Gauss-law-type constraint $\partial_j D^{ij}=0$ imposed by  $A_{i0}$ because it is implied by the chirality constraint. 
The canonical PB relations that one may read off from this action are
\begin{eqnarray}\label{canonicalPBs}
   \left\{ X^m(\boldsymbol{\sigma}), P_n(\boldsymbol{\sigma}'\right\}_{PB} 
   &=& \delta^m_n\,  \delta(\boldsymbol{\sigma}-\boldsymbol{\sigma}')\, , \nonumber \\
    \left\{A_{ij}(\boldsymbol{\sigma}), D^{kl}(\boldsymbol{\sigma}')\right\}_{PB} &=&  \left(\delta_i^k\delta_j^l -\delta_i^l\delta_j^k\right)  \,  \delta(\boldsymbol{\sigma}-\boldsymbol{\sigma}')\, .  
\end{eqnarray}

There is considerable freedom in the choice of the constraint functions ${\cal H}_\mu$; here we shall give them in a different basis to that of \cite{Bergshoeff:1998vx} and with fields 
rescaled\footnote{$A_{ij} \to (\sqrt{2/T}) A_{ij}$ and $D^{ij} \to D^{ij}/\sqrt{2T}$; the notation used here is also different to that of \cite{Bergshoeff:1998vx} and the expression for ${\cal H}_0$ given there has been simplified by expanding out the determinant and then using the ${\cal H}_i=0$ constraint.}  so that the weak-field Hamiltonian density in Monge gauge is independent of the M5-brane tension:
\begin{eqnarray}\label{FullM5cons}
 {\cal H}_0 &=&  \frac12 \left\{\eta^{mn}P_m P_n  +  \frac{T}{2} \left(D_{ij}D^{ij} +B_{ij}B^{ij}\right)  + T^2\det h \right\}\, , \nonumber \\
 {\cal H}_i &=& \partial_i X^m P_m  - V_i\, , 
  \end{eqnarray}
 where $h$ is again the space part of the induced worldvolume metric, and
 \begin{equation}\label{Vi}
 V_i=- \frac14 \varepsilon_{ijklm} D^{ij}B^{lm}  \, . 
\end{equation} 
It should be appreciated that we are using a standard shorthand for which  $D_{ij} = h_{ik}h_{kl} D^{kl}$ and that $\varepsilon_{ijklm}$ is the worldspace alternating invariant tensor density of opposite weight to $\varepsilon^{ijklm}$.   In addition we have the chirality constraint functions
 \begin{equation}
\chi^{ij} = D^{ij} + B^{ij}\, .    
 \end{equation}
Notice that parity flips the relative sign in this expression because $B$ is a pseudo-tensor density of the $O(5)$ rotation group whereas $D$ is a tensor density; a parity flip is needed to recover the results of \cite{Bergshoeff:1998vx} after undoing the rescaling mentioned above. 

Our next task will be to compute the PBs of the constraint functions in order to determine the subset of first-class constraints that are required 
to generate the gauge transformations that are on-shell equivalent to diffeomophisms of the M5-brane worldvolume.

\subsection{Poisson bracket algebra of constraints}

It is convenient to choose  a functional basis  for the constraint functions by defining
\begin{eqnarray}
H_0[\beta] = \int\! d^5\!\sigma \,\beta {\cal H}_0 \, , \qquad H [\boldsymbol{\alpha}] = \int\! d^5\!\sigma \,  \alpha^i {\cal H}_i  \, , 
\end{eqnarray}
where $\beta$ is a scalar inverse-density and $\boldsymbol{\alpha}$ is a 5-vector field, with components $\alpha^i$. We assume that $\beta$ and $\boldsymbol{\alpha}$ are smooth and have compact support, which will allow us to freely integrate by parts without the need to keep surface terms; this is  equivalent to imposing appropriate boundary conditions on the worldvolume fields at spatial infinity. 
Similarly, we define
\begin{equation} 
\chi[\omega] =  \frac12\int\! d^5\!\sigma \, \omega_{ij} \chi^{ij}\, , 
\end{equation}
where $\omega_{ij}$ are the components of a smooth 2-form $\omega$ with compact support.  
We shall see below that the set with functionals $H_0$ and $H$ as elements is first-class, but let us consider first the 
complementary set with functionals $\chi$ as elements; a calculation using the canonical PBs of (\ref{canonicalPBs}) yields
\begin{equation}\label{chichi}
\left\{ \chi[\omega], \chi[\omega']\right\}_{PB} = 2\!\int\! \omega d\omega'\, , 
\end{equation}
where the integral of the 5-form $\omega d\omega'$ is taken over the Euclidean 5-space.  The right hand side is not zero in general, which shows that the 
constraint functions imposing the self-duality condition do not form a first-class set,  but neither do they form a second-class set because the right hand side {\sl is} zero when either $\omega$ 
or $\omega'$ is an exact form. This was to be expected because $\partial_i \chi^{ij} = \partial_i D^{ij}$, which generates the gauge transformation of $A$. 

To compute the remaining PBs of the constraint functionals, it is convenient to begin by establishing that 
\begin{eqnarray}\label{intermed}
\left\{ B^{ij}, (H_0[\beta] + H[\boldsymbol{\alpha}] ) \right\}_{PB} &=& \frac{T}{2} \varepsilon^{ijklm} \partial_k \left(\beta D_{lm}\right) + 3 \partial_k \left(\alpha^{[k} B^{ij]}\right) \nonumber \\
\left\{ D^{ij}, (H_0[\beta] + H[\boldsymbol{\alpha}] )\right\}_{PB} &=& \frac{T}{2} \varepsilon^{ijklm} \partial_k \left(\beta B_{lm}\right)  + 3 \partial_k \left(\alpha^{[k} D^{ij]}\right) \nonumber \\
\left\{X^m, (H_0[\beta] + H[\boldsymbol{\alpha}])\right\}_{PB} &=& \beta P^m + \alpha^i \partial_i X^m \nonumber \\
\left\{P^m,  (H_0[\beta] + H[\boldsymbol{\alpha}])\right\}_{PB} &=& \partial_i \left( \beta\,  {\cal O}^{ij}\partial_j X^m  + \alpha^iP^m\right) 
\end{eqnarray}
where
\begin{equation}
{\cal O}^{ij} = T^2(\det h) h^{ij} + T \left(D^{ik} D^{j\ell} + B^{ik}B^{j\ell}\right)h_{k\ell} \, . 
\end{equation}

It is also convenient to use a different basis by defining 
\begin{equation}
\tilde {\cal H}_i = {\cal H}_i -A_{ij} \partial_k D^{kj} \, , \qquad \tilde H[\boldsymbol{\alpha}] = \int\! d^5\sigma \,  \alpha^i \tilde{\cal H}_i\, . 
\end{equation}
As $\partial_k D^{kj}$ generates the gauge transformations of the  5-space 2-form potential $A$, it has zero PBs with the (gauge-invariant) functionals 
$(H_0,H,\chi)$; this means that PB relations among $(H_0,\tilde H,\chi)$ will be the same as PB relations among $(H_0,H,\chi)$ on the surface in phase space 
determined by the full set of constraints (in any basis). The advantage of this replacement of $H$ by $\tilde H$ is that it leads to  a simpler result for the PB 
relations off this surface. For example,  it is not difficult to establish,  using the intermediate results of (\ref{intermed}),  that 
\begin{equation}
\left\{\tilde H[\boldsymbol{\alpha}], \tilde H[\boldsymbol{\alpha}']\right\}_{PB} = \tilde H \left[ [\boldsymbol{\alpha},\boldsymbol{\alpha}'] \right] \, , 
\end{equation}
where $[\boldsymbol{\alpha},\boldsymbol{\alpha}']$ is the commutator of vector fields; this shows that the constraint functions $\tilde {\cal H}_i$ generate 5-space diffeomorphisms. 
A similar calculation, using the identity
\begin{equation}
\varepsilon^{pijkl} D_{ij}B_{kl} \equiv \det h\, h^{pq} \varepsilon_{qijkl} D^{ij}B^{kl}\, , 
\end{equation}
yields (as should be expected in light of the interpretation just established for $\tilde H$) 
\begin{equation}
 \left\{ \tilde H[\boldsymbol{\alpha}], H_0[\beta]\right\}_{PB} = H_0[{\cal L}_{\boldsymbol{\alpha}}\beta] \, , \qquad 
 \left\{ \tilde H[\boldsymbol{\alpha}], \chi[\omega]\right\}_{PB} =  \chi[{\cal L}_{\boldsymbol{\alpha}}\omega]\, ,
  \end{equation}
 where ${\cal L}_{\boldsymbol{\alpha}}$ is the Lie derivative with respect to $\boldsymbol{\alpha}$. As $\beta$ is a scalar inverse density and $\omega$ a 2-form, we have
 \begin{eqnarray}
 {\cal L}_{\boldsymbol{\alpha}} \beta  &=& \alpha^k\partial_k \beta - (\partial_i\alpha^i)\beta\, ,  \nonumber\\
 \left( {\cal L}_{\boldsymbol{\alpha}}\omega\right)_{ij}  &=& \alpha^k\partial_k \omega_{ij} + 2 (\partial_{[i} \alpha^k) \omega_{j]k} \, . 
 \end{eqnarray}
 So far, these PB relations could have been anticipated from the fact that  the Lagrangian is  an integral over Euclidean 5-space of a scalar density  that is constructed from 5-space 
tensors or tensor-densities.  In addition, one sees easily from (\ref{intermed}) that 
 \begin{equation}\label{hochi}
\left\{ H_0[\beta], \chi[\omega] \right\}_{PB} =  T \chi[\Omega(\beta,\omega)] \, ,  \qquad \Omega_{ij} = \frac12 h_{ip}h_{jq} \varepsilon^{ijkpq} \beta\partial_k \omega_{ij}\, . 
\end{equation}
This has the form required for consistency of the chirality constraint, even though the specific form of $\Omega$ would be hard to guess. 

This leaves the PB relations of the $H_0$ functionals. Using again the intermediate results of (\ref{intermed}), we find that
\begin{eqnarray}\label{h0h0}
\left\{ H_0[\beta], H_0[\beta'] \right\}_{PB} &=& H[\boldsymbol{\alpha}(\beta,\beta')]  \\
&&\!\!\!\! \!\!\!\! \!\!\!\! \!\!\!\! \!\!\!\! \!\!\!\!  -T \int\! d^5\sigma (\beta\partial_i\beta' - \beta'\partial_i\beta)(D^2+B^2)^{ij} V_j\, , \nonumber 
\end{eqnarray}
where
\begin{equation}
\alpha^i(\beta,\beta') = {\cal O}^{ij} (\beta\partial_j\beta' - \beta'\partial_j\beta)\, , 
\end{equation}
and  $(D^2)^{ij} = D^{ik}h_{kl}D^{lj}$ (and similarly for $B^2$). The first term on the right-hand side is expected from known results for branes without worldvolume gauge potentials \cite{Henneaux:1985kr}, 
but the additional term proportional to $T$ is both unexpected and not obviously zero on the surface defined by the constraints. However, it {\it is} zero on this surface; this 
can be seen by using the chirality constraint to replace $D$ by $-B$, which results in 
\begin{equation}
(D^2+B^2)^{ij}V_j \to \frac12 (B^2)^{ij}\varepsilon_{jklpq} B^{kl}B^{pq}  \equiv 0\, , 
\end{equation}
where the following identity  (for free index $i$) has been used:
\begin{equation}
\varepsilon_{jklpq} B_i{}^j B^{kl} B^{pq} \equiv 0\, . 
\end{equation}
Therefore
\begin{equation}\label{simpleho}
\left. \left\{ H_0[\beta], H_0[\beta'] \right\}_{PB} \right|_{\chi=0} =  H[\boldsymbol{\alpha}(\beta,\beta')] \, . 
\end{equation}
This concludes the proof that the constraints corresponding to the functionals $(H_0,\tilde H)$, and hence $(H_0,H)$, form a first-class set.  

Let us recall here that the phase-space action (\ref{M5}) with constraints (\ref{FullM5cons})  is equivalent to the action given in \cite{Bergshoeff:1998vx}, which in turn was derived from the  
worldvolume reparametrization invariant configuration-space action of  \cite{Bandos:1997ui}.  In view of this,  it should not be surprising that the constraints
corresponding to the functionals $(H_0,H)$ form a first-class set; the above results should be seen as a check that no error has crept in along the way.
However, the analysis has thrown up one surprise. 

We know that the chirality constraint is a necessary feature of the full M5-brane action, which includes anticommuting (fermionic) variables, but one might have expected it to be optional 
in the context of the bosonic truncation. This appears not to be the case; if the chirality constraint is replaced by the Gauss-law type constraint $\partial_k D^{kj}=0$, 
then most of the PB relations of the constraint  functions $(H_0,\tilde H)$ are unchanged, but the extra term proportional to $T$ in (\ref{h0h0}) is now non-zero (for non-zero $T$) on the surface defined by the constraints  so the constraints corresponding to these functionals  no longer form a first-class set.  It seems that the bosonic M5-brane action `knows' that its 3-form field-strength must be  
self-dual even in the absence of any worldvolume fermions!

\section{Non-linear 2-form electrodynamics}

Consider now a planar static M5-brane; its worldvolume is a 6D Minkowski spacetime. There are five transverse dimensions into which this Minkowski worldvolume may fluctuate, but we may consistently set to zero these fluctuations, effectively reducing the spacetime dimension from 11 to 6; the M5-brane worldvolume 
is mapped to the 6-dimensional spacetime by the functions $\{ X^\mu(\xi); \mu=0,1,\dots,5\}$.  This is a ``space-filling'' truncation as the M5-brane fills the 5-space; the 
only remaining {\sl physical} fluctuations are those of the 2-form potential on its worldvolume.  The action (\ref{M5}) becomes
\begin{equation}\label{spacefill}
  S = \int\! dt \int\! d^5\sigma \left\{ \dot X^\mu P_\mu + \frac12 \dot A_{ij} D^{ij} - u^\mu {\cal H}_\mu - \frac12\sigma_{ij} \chi^{ij} \right\}  \, , 
\end{equation}
with $\chi^{ij}$ as before but now
\begin{eqnarray}\label{spacefillcon}
 {\cal H}_0 &=&  \frac12 \left\{\eta^{\mu\nu}P_\mu P_\nu  +  \frac{T}{2} \left(D_{ij}D^{ij} +B_{ij}B^{ij}\right)  + T^2\det h \right\}\, , \nonumber \\
 {\cal H}_i &=& \partial_i X^\mu P_\nu - V_i \, . 
 \end{eqnarray}
It remains true that $D_{ij} = h_{ik}h_{kl} D^{kl}$, but now
\begin{equation}
h_{ij}(\xi) = \partial_i X^\mu \partial_j X^\nu \eta_{\mu\nu}\, . 
\end{equation}

This truncated M5-brane action still has a manifest Poincar\'e-invariance, but now in 6D with corresponding Noether charges
\begin{equation}\label{Poinc}
{\cal P}_\mu = \int\! d^5\sigma\,  P_\mu\, , \qquad  {\cal J}^{\mu\nu} = 2\int\! d^5\sigma\, X^{[\mu} P^{\nu]}\, . 
\end{equation}
Using the canonical PB relations, which are now
\begin{eqnarray}\label{canPBs}
   \left\{ X^\mu(\boldsymbol{\sigma}), P_\nu(\boldsymbol{\sigma}'\right\}_{PB} 
   &=& \delta^\mu_\nu\,  \delta(\boldsymbol{\sigma}-\boldsymbol{\sigma}')\, , \nonumber \\
    \left\{A_{ij}(\boldsymbol{\sigma}), D^{kl}(\boldsymbol{\sigma}')\right\}_{PB} &=&  \left(\delta_i^k\delta_j^l -\delta_i^l\delta_j^k\right)  \,  \delta(\boldsymbol{\sigma}-\boldsymbol{\sigma}')\, ,  
\end{eqnarray}
one may verify that the PB algebra of Noether charges is the 6D Poincar\'e algebra. The canonical PB relations may also be use to compute the PBs of the constraint functions. 
This is essentially the same calculation that was detailed earlier, with essentially the same result: the functions ${\cal H}_\mu$ form a  first-class set that generate 6D diffeomorphisms. 

\subsection{Monge gauge}

The diffeomorphism invariance of the action (\ref{spacefill}) allows us to impose the Monge gauge. 
In the current context this is simply an identification of the spacetime coordinates with the worldvolume coordinates: $X^\mu(\xi)= \xi^\mu$. Having made this choice of coordinates we may solve for $P_\mu$:
\begin{eqnarray}\label{solved}
    P^0 &=& \pm T\sqrt{1 + \frac12 T^{-1}(D_{ij}D^{ij} + B_{ij}B^{ij}) + T^{-2} |{\bf V}|^2} \, , \nonumber \\
   {\bf P} &=& {\bf V} \, , 
\end{eqnarray}
where 
${\bf V}$ is the 5-vector density with components $V_i$, as given in (\ref{Vi}).  We should recall here that the Monge-gauge Euclidean 5-space metric is  just the standard Euclidean metric: $h_{ij} =\delta_{ij}$, and that the Monge-gauge Hamiltonian density is $P^0$. The phase-space action in Monge-gauge is therefore
\begin{equation}
 S_{Monge} = \int\! dt \int\! d^5\!\sigma \left\{ \frac12 \dot A_{ij} D^{ij} - P^0
 - \frac12 \sigma_{ij} \chi^{ij}\right\}\, .  
\end{equation}

For weak fields, and assuming $P^0>0$, we have
\begin{equation}
P^0 = T + \frac14 (D_{ij}D^{ij} + B_{ij}B^{ij}) + {\cal O}(T^{-1})\, . 
\end{equation}
We could arrange to cancel the constant $T$ term as we did in the BI case. When this is done the $T\to\infty$ limit can be taken;  this limit is equivalent to the weak-field limit\footnote{This is true only because of the rescaling of fields mentioned previously; the possibility of this rescaling was not appreciated in \cite{Bergshoeff:1998vx} and this accounts for the difficulties found there for the $T\to0$ limit, which we consider in the following section.}. If we also ignore the chirality constraint then elimination of $D^{ij}$ yields the standard, and manifestly Lorentz-invariant, free-field Lagrangian density for 2-form electrodynamics in six-dimensional Minkowski spacetime,  with field equations that propagate three parity-doublets of massless modes. 

The chirality constraint $\chi^{ij}=0$  reduces the number of propagated modes from $6$ to $3$. To see this, we may use this constraint to eliminate $D^{ij}$, after which the weak-field Lagrangian density becomes \cite{Henneaux:1988gg}
\begin{equation}
    {\cal L}_{chiral} = - \frac12\dot A_{ij} B^{ij} -\frac12 B_{ij}B^{ij}\, . 
\end{equation}
The corresponding weak-field equation is
\begin{equation}\label{firstorder}
\dot B^{ij} + \frac12 \varepsilon^{ijklm} \partial_k B_{lm} =0\, , 
\end{equation}
which implies that $\square B^{ij}=0$. There are ten components of $B_{ij}$ but only six are independent (because of the identity $\partial_i B^{ij}=0$) and only 
three of these satisfy the first-order field equation (the other three
satisfy this equation with the opposite relative sign).

\subsection{The strong-field/tensionless limit}

Let us return to the reparametrisation-invariant  action of (\ref{spacefill}) for a ``space-filling'' M5-brane. 
The strong-field limit  at fixed tension $T$ is equivalent to the tensionless limit $T\to0$ at fixed field-strengths, and in this limit
\begin{equation}\label{m5T0cons}
{\cal H}_0 = \frac12 P^2 \, , \qquad {\cal H}_i = \partial_i X^\mu P_\mu - V_i\, . 
\end{equation}
In the functional basis for the constraint functions used previously, the  {\sl non-zero} PB relations are now
\begin{eqnarray}
\left\{\tilde H[\boldsymbol{\alpha}], \tilde H[\boldsymbol{\alpha}']\right\}_{PB} &=& \tilde H \left[ [\boldsymbol{\alpha},\boldsymbol{\alpha}'] \right] \, , \nonumber \\
\left\{ \tilde H[\boldsymbol{\alpha}], H_0[\beta]\right\}_{PB} &=& H_0[{\cal L}_{\boldsymbol{\alpha}}\beta] \, , \nonumber \\
 \left\{ \tilde H[\boldsymbol{\alpha}], \chi[\omega]\right\}_{PB} &=&  \chi[{\cal L}_{\boldsymbol{\alpha}}\omega]\, , 
 \end{eqnarray} 
 together with the PB relation of (\ref{chichi}) for the chirality constraint functionals.  The subalgebra of constraint functions ${\cal H}_\mu$ is now a Lie algebra (as noted for the D3 case in \cite{Mezincescu:2019vxk}). It is also valid  on the full phase space rather than only on the surface defined by the chirality constraint;  this means that the chirality constraint is now optional, in the sense that it can be consistently replaced by the Gauss-law-type constraint imposed by $A_{0i}$.  
 
Notice that the action (\ref{spacefill}) is still Poincar\'e invariant, with the Noether charges of (\ref{Poinc}). In fact, it is conformal invariant. A vector field $k$  is a conformal Killing vector field on  6D Minkowski spacetime if there exists some spacetime scalar function $f_k$ for which 
\begin{equation}
\left({\cal L}_k \eta\right)_{\mu\nu}   = f_k\,  \eta_{\mu\nu}\, ,  
\end{equation}
where ${\cal L}_k\eta$ is the Lie derivative of the Minkowski metric $\eta$ with respect to the vector field $k(X)$.  For any such $k$,  the first-order variation
\begin{equation}
\delta_k X^m=  k^m\, , \qquad \delta_k P_m = - (\partial_m k^n) P_n\, , \qquad \delta_k e =ef_k\, , 
\end{equation}
is an invariance of the action (\ref{spacefill}) with the phase-space functions ${\cal H}_\mu$ of (\ref{m5T0cons}).   
The corresponding Noether charge is 
\begin{equation}\label{confcharge}
Q[k] = \int\! d^5\!\sigma\,  k^\mu P_\mu \, . 
\end{equation}


\subsection{The Monge gauge action and its symmetries} 

In the Monge gauge, $X^\mu(\xi) = \xi^\mu$,  the solution of the constraints ${\cal H}_\mu =0$ at zero tension is
\begin{equation}
P^0 = \pm |{\bf V}|  \, , \qquad {\bf P}={\bf V}\, . 
\end{equation}
Choosing $P^0>0$, the Monge gauge action for the $T\to0$ (equivalently, strong-field) limit is seen to be 
\begin{equation}\label{6Dstrong}
S = \int\! dt \int\! d^5\sigma \left\{ \frac12 \dot A_{ij} D^{ij} - |{\bf V}|
 - \frac12 \sigma_{ij} \chi^{ij}\right\}\, .  
\end{equation}

The field equations of this action are jointly equivalent to 
\begin{equation}
\dot A_{ij} = - B_{ijk}n^k + \sigma_{ij}\, , 
\end{equation}
where $n^i$ are the components of the unit 5-vector field
\begin{equation}
{\bf n} = {\bf V}/|{\bf V}|\, ,  
\end{equation}
and the constraints
\begin{equation}
 D^{ij} + B^{ij} =0\, , \qquad \partial_{[i} \sigma_{jk]}=0\, . 
\end{equation}
Some constraint on the 5-space 2-form $\sigma$ was to be expected because Lagrange multipliers for second-class constraints are
determined by the equations of motion; the fact that only $d\sigma$ is determined was also to be expected because the exact
part of $\sigma$ (in its Hodge decomposition) is what imposes the first-class constraint associated with the abelian 2-form gauge invariance, and Lagrange multipliers for first-class constraints are not determined by the field equations \cite{Dirac:1958sq}.  
The equation for $A$ and the chirality constraint jointly imply the following gauge-invariant field equations:
\begin{equation}\label{DBfieldeqs}
    \dot B^{ij} = -3 \partial_k\left( n^{[k} B^{ij]}\right) \, , \qquad 
    \dot D^{ij} = -3 \partial_k\left( n^{[k} D^{ij]}\right)\, . 
\end{equation}
Notice that these equations imply 
\begin{equation}
\dot\chi^{ij} = 3 \partial_k \left(n^{[i}\chi^{jk]}\right)\, , 
\end{equation}
which shows that the chirality constraint effects a consistent truncation of the equations that we would have without this
constraint. 

This action (\ref{6Dstrong}) is manifestly invariant under 5-space rotations but Lorentz invariance is no longer manifest. 
Nevertheless, it {\sl is} Lorentz invariant. The Noether charges are unchanged by gauge fixing except that we
must use the gauge-fixed expressions when evaluating them or taking Poisson brackets, which must now be computed 
using the canonical PBs of the gauge-fixed action:
\begin{equation}
 \left\{A_{ij}(\boldsymbol{\sigma}), D^{kl}(\boldsymbol{\sigma}')\right\}_{PB} = 
 \left(\delta_i^k\delta_j^l -\delta_i^l\delta_j^k\right)  \,  \delta(\boldsymbol{\sigma}-\boldsymbol{\sigma}')\, . 
\end{equation}
For example, the Monge-gauge charges for time and space translations are
\begin{equation}
{\cal P}^0 = \int \! d^5\sigma \, |{\bf V}|\, , \qquad \boldsymbol{{\cal P}} = \int\! d^5\sigma \, {\bf V}\, . 
\end{equation}
It is instructive to verify that these quantities are conserved as a consequence of the equations of motion. This can be
done even without the use of the chirality constraint but in this case one must use instead the Gauss-law-type constraint
$\partial_i D^{ij}=0$. Using the identity
\begin{equation}
    \frac12 (\partial_p n^j) \varepsilon_{ijklm}\left(D^{pk}B^{lm} + B^{pk}D^{lm}\right) \equiv 
    (\partial_i n^j)V_j - (\partial_j n^j) V_i\, .  
\end{equation}
and the fact that ${\bf V}= {\bf P}$, one finds that
\begin{equation}\label{dotP}
\dot P_i = - \partial_j (n^jP_i)  \quad \left(\Rightarrow \quad 
\dot P^0 = - \boldsymbol{\nabla} \cdot {\bf P}\right)\, , 
\end{equation}
and hence $\dot{\cal P}_\mu=0$ for appropriate boundary conditions.  

More generally, the Monge-gauge expression for the Noether charge (\ref{confcharge}) associated to any vector field $k$ is
\begin{equation}
Q[k] = \int\! d^5\sigma \, k^\mu(t,\boldsymbol{\sigma}) P_\mu\, , 
\end{equation}
and so 
\begin{eqnarray}
    \dot Q[k] &=& \int\! d^5\sigma \,\left\{ \dot k^0 P_0 + \dot k^iP_i + k^0 \dot P_0 + k^i \dot P_i\right\}\nonumber \\
  &=&   \int\! d^5\sigma \,\left\{ \dot k^0 P_0 + (\dot {\bf k} - \boldsymbol{\nabla} k^0)\cdot {\bf P} + (\partial_jk^i) n^jn_iP^0\right\}\nonumber \\
  && \qquad + \ \int\! d^5\sigma \, \boldsymbol{\nabla} \cdot \left[ \left(k^0- {\bf n}\cdot{\bf k}\right) {\bf P}\right]\, , 
    \end{eqnarray}
where  the second equality involves the use of the equations (\ref{dotP}), an integration by parts, and use of the relation ${\bf P}= {\bf n}P^0$.  Next, we observe that $k$ generates a conformal isometry of the 6D Minkowski metric $\eta$ if 
${\cal L}_k\eta = f_k \eta$ for some function $f_k$; this is equivalent to the equations
\begin{equation}
    \dot k^0=f_k\, , \qquad \dot {\bf k} - \boldsymbol{\nabla} k^0 =0\, , \qquad \partial_{(i} k_{j)} = f_k\delta_{ij}\, , 
\end{equation}
from which we deduce that $\dot Q[k]=0$ for approriate boundary conditions. This establishes conformal invariance of the action (\ref{6Dstrong}). 

\subsection{Lorentz invariance and the stress tensor}

The Lorentz boost generator is
\begin{equation}
    {\bf L} = t \boldsymbol{{\cal P}} - \int\! d^5\!\sigma\, \boldsymbol{\sigma}P^0\, . 
\end{equation}
In Monge gauge we find, for constant uniform 5-vector parameter ${\bf w}$, that 
\begin{equation}\label{LorentzA}
\left\{A_{ij},  {\bf w}\cdot {\bf L}\right\}_{PB} = \frac12 \varepsilon_{ijklm}N^k B^{lm}\, ,
\end{equation}
where 
\begin{equation}\label{5vecN}
{\bf N} = t{\bf w} - ({\bf w}\cdot\boldsymbol{\sigma}) {\bf n}\, \, . 
\end{equation}
This implies that
\begin{equation}
    \left\{B^{ij},  {\bf w}\cdot {\bf L}\right\}_{PB} = 3\partial_k\left(N^{[k} B^{ij]}\right) \, . 
\end{equation}
A separate PB calculation yields
\begin{equation}
\left\{D^{ij},  {\bf w}\cdot {\bf L}\right\}_{PB} = 3\partial_k\left(N^{[k} D^{ij]}\right)\, . 
\end{equation}
A consequence of the above results is that 
\begin{equation}
    \left\{ \chi^{ij},  {\bf w}\cdot {\bf L}\right\}_{PB} = 3\partial_k\left(N^{[k} \chi^{ij]}\right)\, ,  
\end{equation}
which shows that the chirality constraint is Lorentz invariant. Another consequence is 
\begin{eqnarray}\label{Ltrans}
\left\{ {\bf V}, {\bf w}\cdot {\bf L}\right\}_{PB} &=& - {\bf w} |{\bf V}| + \partial_k\left(N^k {\bf V}\right)\, , \nonumber \\
\left\{ |{\bf V}|, {\bf w}\cdot {\bf L}\right\}_{PB} &=& - ({\bf w}\cdot{\bf n}) |{\bf V}| + \partial_k\left(N^k |{\bf V}|\right)\, .  
\end{eqnarray}
Of course, the first of these equations implies the second. 

Recall that $P_\mu$ in Monge gauge has components $(-|{\bf V}|, {\bf V})$. However, these are {\it not} the components of
a Lorentz co-vector field, despite the notation\footnote{Their 5-space integrals are the components of a Lorentz 6-vector, however, so the notation has some justification.}. Let us define a new set of components:
\begin{equation}
\hat P_\mu = P_\mu/\sqrt{P^0} \, . 
\end{equation}
In Monge gauge
\begin{equation}
\hat P_0 = - \sqrt{|{\bf V}|} \, , \qquad \hat {\bf P} = {\bf V}/\sqrt{|V|}\, . 
\end{equation}
These {\it are} the components of a Lorentz co-vector field.  A direct calculation using (\ref{Ltrans}) yields 
\begin{eqnarray}\label{PBhats}
\left\{\hat {\bf P}, {\bf w}\cdot {\bf L}\right\}_{PB} &=& N^k\partial_k \hat{\bf P} + 
\hat P_0 \left[ {\bf w} + \frac12{\bf n} ({\bf w}\cdot\boldsymbol{\sigma}) (\boldsymbol{\nabla}\cdot {\bf n}) \right]\, , \\
\left\{\hat P_0, {\bf w}\cdot {\bf L}\right\}_{PB} &=& N^k\partial_k \hat P_0 + 
\hat P_0 \left[ ({\bf w}\cdot {\bf n}) + \frac12 ({\bf w}\cdot\boldsymbol{\sigma}) (\boldsymbol{\nabla}\cdot {\bf n})\right]\, . \nonumber
\end{eqnarray} 
Again, the first of these equations implies the second. 

This result for the Lorentz transformation of $\hat P_\mu$ may be compared with its Lie derivative with respect to the vector field $\zeta$ with components
\begin{equation}
  \zeta^0   = {\bf w}\cdot {\boldsymbol{\sigma}}\, , \qquad \boldsymbol{\zeta} = t {\bf w}\, . 
\end{equation}
Using the equations (\ref{dotP}) to replace time-derivatives by space derivatives, one may calculate this Lie derivative  from its definition: 
\begin{equation}
    {\cal L}_\zeta \hat P_\mu := \zeta^\nu\partial_\nu \hat P_\mu + (\partial_\mu\zeta^\nu)\hat P_\nu \, . 
\end{equation}
Comparing the result with (\ref{PBhats}), one sees that 
\begin{equation}
\left\{\hat P_\mu, {\bf w}\cdot {\bf L}\right\}_{PB} = ({\cal L}_\zeta \hat P)_\mu\, . 
\end{equation}
To see that this {\it is} a Lorentz transformation we observe (i) that to first order in the 5-vector parameter ${\bf w}$,
\begin{equation}
    t \to t' = t + ({\bf w}\cdot\boldsymbol{\sigma}) \, , \qquad 
    \boldsymbol{\sigma} \to \boldsymbol{\sigma} = \boldsymbol{\sigma} + t {\bf w}\, , 
\end{equation}
is a Lorentz boost transformation of the spacetime coordinates, and (ii) that for any scalar field $\Phi$ we have, 
again to first order in ${\bf w}$, 
\begin{equation}
\Phi(t', \boldsymbol{\sigma}') = \Phi(t,\boldsymbol{\sigma}) + \left[{\cal L}_\zeta \Phi\right](t,\boldsymbol{\sigma})\, . 
\end{equation}
Similarly, the first-order variation of any tensor or tensor-density field under a Lorentz boost with parameter ${\bf w}$ is given by 
its Lie derivative with respect to $\zeta$. 

Since $\hat P_\mu$ is a Lorentz 6-vector, it follows that 
\begin{equation}\label{stress1}
T_{\mu\nu} = \hat P^\mu \hat P^\nu = (P^0)^{-1}P_\mu P_\nu
\end{equation}
is a symmetric tensor of the Lorentz group. It is also `conserved' in the sense that 
\begin{equation}
    \partial^\mu T_{\mu\nu} = 0\, . 
\end{equation}
To check this we observe that we already know from (\ref{dotP}) that $\partial^\mu P_\mu=0$, so that 
\begin{equation}
\partial^\mu T_{\mu\nu} = P^\mu \partial_\mu (P_\nu/P^0)\, . 
\end{equation}
The right hand side is trivially zero for $\nu=0$, so (taking into account that ${\bf P}/P^0= {\bf n}$) we 
see that what needs to be checked is that 
\begin{equation}
\dot n^i +  {\bf n}\cdot \boldsymbol{\nabla}n^i= 0\, ,  
\end{equation}
but this is a consequence of the equations of (\ref{dotP}).

The expression (\ref{stress1}) for the stress tensor may be rewritten in terms of the energy density $P^0=|{\bf V}|$ 
and a null 6-vector $n$ as
\begin{equation}
T_{\mu\nu} = n_\mu n_\nu |{\bf V}| \, ,   \qquad n_\mu = (-1,{\bf n})\, , 
\end{equation}
Since $n$ is null this stress tensor is traceless, as it must be given the conformal invariance of the field equations. 
This is the stress tensor of a null fluid, as found for BBE in \cite{BialynickiBirula:1992qj} and in agreement with \cite{Gibbons:2000ck}. 

\subsection{Simplifying the action} 

So far we have maintained the chirality constraint as one  imposed by a Lagrange multiplier in the action. 
This has the advantage that the symplectic form defined by the phase-space action is in standard Darboux form, from which 
we can easily read off the PBs, which we have been using extensively in the above discussion of symmetries of the action. 
However, now that we have dealt with this  topic it is convenient to use the chirality constraint to 
eliminate $D$ from the action; we simply replace it by $-B$. We then have 
\begin{equation}
    V_i \to  \frac14\varepsilon_{ijklm} B^{jk}B^{lm} \equiv (B\wedge B)_i\, , 
\end{equation}
and the action (\ref{6Dstrong}) becomes a functional of the space 2-form $A$ alone:
\begin{equation}\label{altstrong}
S[A] = \int\! dt \int\! d^5\sigma \left\{-\frac12\dot A_{ij} B^{ij} - |B\wedge B| \right\} \, . 
\end{equation}
This is the action advertised in the Introduction. Its field equations are, as could be expected from (\ref{DBfieldeqs}),
\begin{equation}\label{dotB}
\dot B^{ij} = 3\partial_k \left(n^{[i}\,  B^{jk]} \right)\, , 
\end{equation}
where the direction of the unit vector field  ${\bf n}$ is now determined by $B\wedge B$. These are non-linear equations, analogous to the BBE equations (\ref{BBeofm}); as in that case, the non-linearity is due to the dependence of the unit $5$-vector field ${\bf n}$ on $B$. 

It is instructive to verify the Lorentz invariance of this simplified action. From (\ref{LorentzA}) we see that the first-order 
Lorentz transformation of $A$ is 
\begin{equation}
    \delta_{\bf v} A_{ij} = \frac12 \varepsilon_{ijklm}N^k B^{lm}\, . 
\end{equation}
where ${\bf N}$ is given in (\ref{5vecN}). This implies\footnote{Here one must use the fact that $(B\wedge B)_i = n_i|B\wedge B|$ and the identity ${\bf n}\cdot d{\bf n}\equiv 0$.} 
\begin{equation}
\delta_{\bf v} |B\wedge B| = - ({\bf n}\cdot {\bf v}) |B\wedge B| + {\rm total\ space\ derivative}
\end{equation}
The Lorentz variation of the Hamiltonian is therefore {\sl not} zero but we should not expect it to be zero because the 
transformation of $A$ involves an explicit time-dependence through its dependence on ${\bf N}$, and this will produce 
a variation of the geometric term in the action. Specifically, one finds that 
\begin{equation}
\delta_{\bf v} \left(- \dot A_{ij} B^{ij} \right) = - \, ({\bf n}\cdot {\bf v})|B\wedge B| + {\rm total\ time \ derivative} \, , 
\end{equation}
where the second line makes use of the identity ${\bf n}\cdot \dot{\bf n}\equiv0$. 
The action $S[A]$ is therefore Lorentz invariant, despite appearances.

We also learn something else from this check of Lorentz invariance. It might be thought that we could generalise
the action $S[A]$ to 
\begin{equation}
    S_\lambda[A] = \int\! dt \int\! d^5\sigma \left\{-\frac12\dot A_{ij} B^{ij} - \lambda |B\wedge B| \right\} 
\end{equation}
for arbitrary constant $\lambda$. However, this generalised action is Lorentz invariant {\sl only} for $\lambda=1$ (given our assumption of a positive Hamiltonian density). Also, it is not possible to introduce $\lambda$ by rescaling 
$A$ since both terms in the action (the geometrical term and the Hamiltonian) are homogeneous of second degree in $A$.  Despite this fact, the Hamiltonian is {\sl not} quadratic in $A$; if it were, $S[A]$ would be a free-field action, with free-field equations, 
but the equations are not free-field equations.

\section{IR limit of an M5-brane in AdS$_7\times S^4$}

So far we have considered the M5-brane, or some truncation of its dynamics, in the 
11D Minkowski vacuum of 11D supergravity. Now we turn to the M5-brane in the 
AdS${}_7\times S^4$ vacuum. For an $S^4$ of radius $R$, the  AdS${}_7\times S^4$ metric is
\begin{equation}
ds^2_{11} = \left(\frac{r}{R}\right) dx^\mu dx^\nu \eta_{\mu\nu}   + \left(\frac{R}{r}\right)^2 \left(dr^2 + r^2 d\Omega_4^2\right)\, , 
\end{equation}
where $r$ is the radial coordinate for spherical polar coordinates on $\mathbb{E}^5$ and $d\Omega_4^2$ is the $SO(5)$-invariant unit metric on $S^4$. If we set
\begin{equation}
r= 4R^3/z^2
\end{equation}
then the metric becomes 
\begin{equation}
    \left(\frac{2R}{z}\right)^2 \left[ dx^\mu dx^\nu \eta_{\mu\nu} + dz^2 + \left(\frac{z}{2}\right)^2 d\Omega_4^2\right]\, , 
\end{equation}
where $z$ is now an inverse-square radial coordinate. The boundary of AdS$_7$ is at $z=0$
whereas the Killing horizon of our Poincar\'e patch coordinates is at $z=\infty$. 

We now consider a static planar M5-brane in this background, at fixed $z$ and fixed position on the 4-sphere, so that its worldvolume is coincident with the 6D Minkowski space with coordinates $x^\mu$. This is a solution of the M5-brane equations of motion and we wish to consider fluctuations about it. The induced metric on this fluctuating M5-brane is
\begin{equation}
    h_{ij} = \left(\frac{2R}{z}\right)^2\left[\partial_i x^\mu \partial_j x^\nu\eta_{\mu\nu} + \partial_i z\partial_j z + (z/2)^2 \partial_i\psi^I\partial_j\psi^J \bar g_{IJ}\right]\, , 
\end{equation}
where $\bar g_{ij}$ is the metric on the unit 4-sphere in angular coordinates $\{\psi^I; I=1,2,3,4\}$. 
If we now use this result for the induced metric in the constraint functions (\ref{FullM5cons})
for the M5-brane phase-space action, and rescale $u^0 \to \tilde u^0$ such that 
$u^0{\cal H}_0 = \tilde u^0\tilde {\cal H}_0$ for 
\begin{equation}
\tilde {\cal H}_0 = \left(\frac{2R}{z}\right)^2 {\cal H}_0\, , 
\end{equation}
then we find that 
\begin{eqnarray}
\tilde {\cal H}_0 &=& \frac12 \left\{ \eta^{\mu\nu}p_\mu p_\nu + p_z^2 + 
\frac{4L^2}{z^2} + {\cal O}( TR^6/z^6) \right\}\, ,  \nonumber \\
{\cal H}_i &=& \partial_i x^\mu p_\mu + \partial_i z p_z + \partial_i\psi^I p_I 
- (B\wedge B)_i\, , 
\end{eqnarray}
where 
\begin{equation}
L^2 = \bar g^{IJ}p_i p_J\, , 
\end{equation} 
which is the square of the $S^4$ angular momentum, and we have used the chirality constraint to set $D=-B$ in the equation for ${\cal H}_i$. 

However, the constraint functions (\ref{FullM5cons}) require some $T$-dependent modifications when the 
11D supergravity 3-form potential $C^{(3)}$ is non-zero, as it is for the AdS${}_7\times S^4$ vacuum since $dC^{(3)}$ is proportional to the volume 4-form on $S^4$ \cite{Pilch:1984xy}. In addition, there is a
coupling, with coefficient $T$, to the dual 6-form $C^{(6)}$ that is defined for solutions of the 11D 
supergravity field equations; it takes the following form for the AdS$_7\times S^4$ solution \cite{Claus:1998mw}:
\begin{equation}
C^{(6)} \propto \left(\frac{2R}{z}\right)^6 dx^0dx^1\cdots dx^5\, . 
\end{equation}
The coupling to $C^{(6)}$ therefore contributes a further ${\cal O}(TR^6/z^6)$ term. 

As these modifications all come with a factor of $T$ they are not relevant to the $T\to0$ limit, but now we
can consider a different limit in which $z\to\infty$. Recall that as $z\to 0$ the M5-brane moves to the 6D Minkowski boundary of AdS$_7$, where its dynamics becomes that of a free field theory. In the context of the AdS/CFT correspondence, this is the UV limit of the M5-brane dynamics. The $z\to\infty$ limit is an IR limit in which the brane moves to the null Killing horizon (which is the boundary of the Poincar\'e patch covered by the coordinates that we chose for the AdS$_7$ metric).  All ${\cal O}(TR^6/z^6)$ terms can still be ignored,  now because they involve inverse powers of $z$, 
but this still leaves modifications arising from couplings to $C^{(3)}$, which takes the form
\begin{equation}
C^{(3)} \propto R^3 c_{IJK}(\{\psi\}) d\psi^I d\psi^J d\psi^K\, .
\end{equation}

All modifications due to the non-zero background form fields, in particular those due to $C^{(3}$,  can be viewed (by field redefinitions) as modifications to the phase-space constraints {\sl only}. One such modification is $B \to B -T {\cal C}^{(3)}$, where $B=dA$ and ${\cal C}^{(3)}$ is the pullback of $C^3$ to the M5-brane. Another is the replacement $P_m \to P_m - TC_m$, where $C_m$ is the 5-space Hodge-dual of the 5-form obtained by contraction with the vector field $\partial_m$ of a 6-form consisting of a linear combination of  $C^{(3)} \wedge C^{(3)}$ and $C^{(3)}\wedge B$ \cite{Bergshoeff:1998vx}. In each case\footnote{The configuration space action of \cite{Bandos:1997ui} includes the worldvolume integral of ${\cal C}^{(3)}\wedge B$, which contributes an additional term involving $\dot A$, but this is cancelled by another such term from elsewhere in the configuration space action, so no redefinition of $D$ is needed \cite{Bergshoeff:1998vx}.} there is a factor of ${\cal C}^{(3)}$. These terms are also irrelevant in the $z\to\infty$ limit, provided that the limit is taken keeping
the momentum variables $p_I$, and hence $L^2$, finite. The $L^2/z^2$ term then drops out of $\tilde {\cal H}_0$ and the $p_I$ variables in the action become Lagrange multipliers for the constraints
\begin{equation}
    \dot\psi^I - u^i\partial_i \psi^I =0\, .  
\end{equation}
These equations should hold for all possible $u^i$, which are generically non-zero and vary with solutions of the equations of motion. The position of the M5-brane worldvolume on $S^4$ is therefore fixed (independent of worldvolume coordinates) in the $z\to\infty$ limit.  

To conclude, the $z\to\infty$ limit leads to the action 
\begin{equation}
    S^{(IR)}_{M5}= \int\! dt\int\! d^5\!\sigma \left\{ \dot x^\mu p_\mu + \dot z p_z  - \frac12\dot A_{ij}B^{ij} - \tilde u^0 \tilde{\cal H}_0 - u^i{\cal H}_i \right\}
\end{equation}
where
\begin{equation}
    \tilde {\cal H}_0 = \eta^{\mu\nu} p_\mu p_\nu + p_z^2 \, , \qquad 
    {\cal H}_i = \partial_i x^\mu p_\mu + \partial_i z p_z - (B\wedge B)_i\, . 
\end{equation}
Choosing the Monge gauge and solving the constraints yields the equivalent action 
\begin{equation}
    S= \int\! dt\int\! d^5\!\sigma \left\{\dot z p_z - \frac12\dot A_{ij} B^{ij} - {\cal H} \right\}
\end{equation}
where
\begin{equation}
{\cal H} = \sqrt{\delta^{ij} p_i p_j + p_z^2} \, , \qquad p_i= (B\wedge B)_i - \partial_i z p_z\, . 
\end{equation}

A consistent truncation of this action is to set $dz=0$ and $p_z=0$, in which case we recover the 
chiral 2-form electrodynamics with the action S[A] of (\ref{strongfield}). This derivation of it allows
us to interpret it as the dynamics of a chiral 2-form on an M5-brane at the null Killing horizon of AdS$_7$.

\section{Discussion}

It has been shown here that the interacting conformal-invariant Bialynicki-Birula electrodynamics (BBE) \cite{BialynickiBirula:1984tx}
has an analog in six dimensions but for a 2-form potential,  which is subject to a chirality condition that halves the number of
degrees of freedom. Just as BBE is a strong-field limit of the non-linear (and non-conformal)  Born Infeld electrodynamics, which is a truncation of the
Dirac-Born-Infeld action for a D3-brane, the new conformal chiral 2-form electrodynamics is a strong-field limit of the non-linear (and non-conformal)
2-form electrodynamics that survives a similar truncation of the analogous action for an M5-brane.  In both cases this strong-field limit can also be taken
{\it prior} to the truncation, in two different ways.

One way is to take the tensionless limit, either of a D3-brane in the 10D Minkowski vacuum of IIB supergravity (as done in \cite{Mezincescu:2019vxk},  where
it leads to a generalization of BBE to include scalar fields)  or of an M5-brane in the 11D Minkowski vacuum of 11D  supergravity, which  yields a similar 
generalization (not explored here) of the 6D chiral 2-form electrodynamics. These limits must be taken within a phase-space formulation  of the action, but these are
known from earlier work in  \cite{Bergshoeff:1998ha}  (D-branes) and \cite{Bergshoeff:1998vx} (M5-brane); in the latter work  the tensionless limit of the M5-brane 
action was also considered but with inconclusive results; the problematic features of this limit that were noted there have been  resolved here. Essentially,  one 
should choose definitions of the worldvolume fields for which the tension drops out in the weak-field limit; the tensionless limit then 
becomes possible and is  equivalent to a strong-field limit. 

The other way to arrive at both BBE and the new 6D chiral 2-form electrodynamics is to consider, respectively,  a D3-brane in the AdS$_5\times S^5$ 
vacuum of 10D IIB supergravity and an M5-brane in the AdS$_7\times S^4$ vacuum of 11D supergravity.  In Poincar\'e patch coordinates for AdS
an infinite static planar brane, with Minkowski worldvolume, will solve the brane equations of motion if it is placed at a fixed radial distance from the 
Killing horizon in these coordinates. Moving the brane to larger distances corresponds to a renormalization group flow towards the ultra-violet limit of 
the brane dynamics (which is the free conformal theory of the weak-coupling limit)  whereas moving it closer to the horizon corresponds to a flow towards
the IR.  Again, the limit cannot be taken in the configuration-space form of the brane action, but it can be taken in the phase-space form of the action.

As shown for the D3-brane in \cite{Mezincescu:2019vxk} and here for the M5-brane, this IR limit describes a brane at the AdS Killing horizon with all transverse
fluctuations frozen except for fluctuations into the AdS bulk, determined by a single scalar. This is generic, and would apply to the M2-brane in AdS$_4\times S^7$ 
but for the D3-brane and M5-brane there is the additional dynamics of the worldvolume  albelian gauge potential, a 1-form for D3 and a 2-form for M5. A consistent 
truncation  to this form-field dynamics yields precisely BBE in the D3 case  and the new conformal chiral 2-form electrodynamics in the M5 case.

Whereas a tensionless limit of the D3-brane or M5-brane is a rather artificial one within the String/M-theory context for which these branes have physical relevance,
the IR limit just desribed is a natural one to consider. The  `AdS$\times S$' background is just a low-energy description of the effect of a large number, $N+1$ say, 
of  parallel coincident branes;  from an AdS/CFT perspective the `AdS$\times S$' background is the AdS bulk description of the ground state of the lR dynamics 
of these branes. If one of the $N+1$ branes is separated from the remaining $N$ but remains parallel to them then it will appear as a probe brane with a worldvolume
coincident with one of the 6D Minkowski slices of the Poincar\'e patch of the ADS spacetime (with a slightly increased constant radius of curvature). The IR limit
thus corresponds to the return to the fold of a lone and isolated brane; it is therefore plausible that its IR dynamics will contain some information about the 
IR dynamics of the collection of $N+1$ branes. In the M5-case this is the ``mysterious'' (2,0)-supersymmetry 6D CFT. 

This brings us to the question of whether the  conformal chiral 2-form 6D electrodynamics described here can be incorporated into a supersymmetric extension, 
in particular a ($2,0)$-supersymmetric extension.  One could attempt to answer this question directly but it could also be addressed by returning to the phase-space action for 
the M5-brane but now including all fermions; this action was given in \cite{Bergshoeff:1998vx}. The weak-field limit of the configuration space M5-brane action of 
\cite{Bandos:1997ui} was worked out in detail in \cite{Claus:1998mw}; not surprisingly, this limit reduces the M5-brane dynamics to that of a free $(2,0)$ 6D supermultiplet. 
It is therefore plausible that the strong-field limit of  the full phase-space action of the M5-brane will be a $(2,0)$-supersmmetric extension of the simple chiral 2-form electrodynamics 
presented here, but any detailed verification of this is likely to require  significantly more effort than the author has exerted to obtain the results reported here. 

\vfill
\section*{Acknowledgements}
This work was partially supported by STFC consolidated grant ST/L000385/1. It was an outgrowth of 
work with Luca Mezincescu, to whom the author is grateful for helpful comments; thanks are also due to
Gary Gibbons for bringing to the author's attention his earlier work with Peter West.



\begin{thebibliography}{10}

\bibitem{BialynickiBirula:1984tx}
  I.~Bialynicki-Birula,
  ``Nonlinear Electrodynamics: Variations On A Theme By Born And Infeld,''
  In {\sl Quantum Theory Of Particles and Fields}, eds. B. Jancewicz and 
J.  Lukierski,  (World Scientific, 1983) pp  31-48. 

 
 
\bibitem{Mezincescu:2019vxk}
  L.~Mezincescu and P.~K.~Townsend,
  ``DBI in the IR,''
  J.\ Phys.\ A {\bf 53} (2020) no.4,  044002
  [arXiv:1907.06036 [hep-th]].

  
  
\bibitem{Duff:2018goj}
  M.~J.~Duff,
  ``Thirty years of Erice on the brane,''
  arXiv:1812.11658.
  
\bibitem{Townsend:1995af}
  P.~K.~Townsend,
  ``D-branes from M-branes,''
  Phys.\ Lett.\ B {\bf 373} (1996) 68
  [hep-th/9512062].
  
 
\bibitem{Strominger:1995ac}
  A.~Strominger,
  ``Open p-branes,''
  Phys.\ Lett.\ B {\bf 383} (1996) 44
  [hep-th/9512059].
  
\bibitem{Witten:1995zh}
  E.~Witten,
  ``Some comments on string dynamics,''
  hep-th/9507121.

\bibitem{Arvidsson:2004xa}
  P.~Arvidsson, E.~Flink and M.~Henningson,
  ``The (2,0) supersymmetric theory of tensor multiplets and selfdual strings in six-dimensions,''
  JHEP {\bf 0405} (2004) 048
  [hep-th/0402187].
  
  
\bibitem{Bandos:1997ui}
  I.~A.~Bandos, K.~Lechner, A.~Nurmagambetov, P.~Pasti, D.~P.~Sorokin and M.~Tonin,
  ``Covariant action for the superfive-brane of M theory,''
  Phys.\ Rev.\ Lett.\  {\bf 78} (1997) 4332
  [hep-th/9701149].
  
\bibitem{Bergshoeff:1998vx}
  E.~Bergshoeff, D.~P.~Sorokin and P.~K.~Townsend,
  ``The M5-brane Hamiltonian,''
  Nucl.\ Phys.\ B {\bf 533} (1998) 303
  [hep-th/9805065].
  
\bibitem{Gibbons:2000ck}
  G.~W.~Gibbons and P.~C.~West,
  ``The Metric and strong coupling limit of the M5-brane,''
  J.\ Math.\ Phys.\  {\bf 42} (2001) 3188
  [hep-th/0011149].

  
\bibitem{Bergshoeff:1998ha}
  E.~Bergshoeff and P.~K.~Townsend,
  ``Super D-branes revisited,''
  Nucl.\ Phys.\ B {\bf 531} (1998) 226
  [hep-th/9804011].
  
\bibitem{Henneaux:1985kr}
  M.~Henneaux,
  ``Hamiltonian Form of the Path Integral for Theories with a Gauge Freedom,''
  Phys.\ Rept.\  {\bf 126} (1985) 1.

  
\bibitem{Henneaux:1988gg}
  M.~Henneaux and C.~Teitelboim,
  ``Dynamics of Chiral (Selfdual) $P$ Forms,''
  Phys.\ Lett.\ B {\bf 206} (1988) 650.
  
  
\bibitem{Dirac:1958sq}
  P.~A.~M.~Dirac,
  ``Generalized Hamiltonian dynamics,''
  Proc.\ Roy.\ Soc.\ Lond.\ A {\bf 246} (1958) 326.
  
\bibitem{BialynickiBirula:1992qj}
  I.~Bialynicki-Birula,
  ``Field theory of photon dust,''
  Acta Phys.\ Polon.\ B {\bf 23} (1992) 553.


  
 

\bibitem{Pilch:1984xy}
  K.~Pilch, P.~van Nieuwenhuizen and P.~K.~Townsend,
  ``Compactification of $d=11$ Supergravity on S(4) (Or 11 = 7 + 4, Too),''
  Nucl.\ Phys.\ B {\bf 242} (1984) 377.
  
\bibitem{Claus:1998mw}
  P.~Claus, R.~Kallosh, J.~Kumar, P.~K.~Townsend and A.~Van Proeyen,
  ``Conformal theory of M2, D3, M5 and D1-branes + D5-branes,''
  JHEP {\bf 9806} (1998) 004
  [hep-th/9801206].



\end{thebibliography}
\end{document}